\documentclass[12pt]{article}
\usepackage{a4p}
\usepackage{cite}
\usepackage{graphicx}
\usepackage{amsmath} 

\newcommand{\Z}  {\ensuremath{\mathrm{Z}^0}}
\newcommand{\T}  {\ensuremath{\tau}}
\newcommand{\tl} {{$\tau$} lepton}
\newcommand{\tls}{{$\tau$} leptons}
\newcommand{\td} {{$\tau$} decay}
\newcommand{\tds}{{$\tau$} decays}
\newcommand{\tm} {{$\tau$} mass}
\newcommand{\tp} {{$\tau$} pair}
\newcommand{\tps}{{$\tau$} pairs}
\newcommand{\ma} {\ensuremath{m_{\mathrm{cn}}^\ast}}
\newcommand{\mc} {\ensuremath{m_{\mathrm{ev}}^\ast}}
\newcommand{\mstar} {\ensuremath{m^\ast}}
\newcommand{\nut}{$\nu_\tau$}
\newcommand{\Eh} {\ensuremath{E_{\mathrm{h}}}}
\newcommand{\ph} {\ensuremath{p_{\mathrm{h}}}}
\newcommand{\vph}{\ensuremath{\vec{p}_{\mathrm{h}}}}
\newcommand{\mh} {\ensuremath{m_{\mathrm{h}}}}
\newcommand{\mt} {\ensuremath{m_\tau}}

\newcommand{\PSI}   {\ensuremath{\psi}}
\newcommand{\PSIp}  {\ensuremath{\psi_+}}
\newcommand{\PSIm}  {\ensuremath{\psi_-}}

\newcommand{\KO}     {\ensuremath{\mathrm{K}^0}}

\newcommand{\taup}   {\ensuremath{\tau \rightarrow \pi^\pm\,\nu_\tau}}
\newcommand{\taupn}  {\ensuremath{\tau \rightarrow \pi^\pm\,\pi^0\,\nu_\tau}}
\newcommand{\taupnn}{\ensuremath{\tau \rightarrow \pi^\pm\,2\,\pi^0\,\nu_\tau}}

\newcommand{\tauppp} {\ensuremath{\tau \rightarrow 3\,\pi^\pm\,\nu_\tau}}
\newcommand{\taupppn}{\ensuremath{\tau \rightarrow 3\,\pi^\pm\,\pi^0\,\nu_\tau}}
\newcommand{\taupk}  {\ensuremath{\tau \rightarrow \pi^\pm\,\KO\;\nu_\tau}}
\newcommand{\ee}     {\ensuremath{\mathrm{e}^+\,\mathrm{e}^-}}

\newcommand{\ppi}    {\ensuremath{\pi^\pm}}
\newcommand{\pni}    {\ensuremath{\pi^0}}
\newcommand{\dd}     {\ensuremath{~\leftrightarrow~}}
\newcommand{\emp}    {\multicolumn{1}{c|}{}}

\def\etal      {{\it et~al.}}

\def\mtau {1775.1}                             
\def\estat{1.6}                                
\def\esys      {1.0}                           
\def\estatMC   {0.4}                           
\def\esysbeam  {40}                            
\def\esysmnu   {18.2}                          
\def\calibCT   {$0.1\:\%$}                     
\def\calibEB   {$0.3\:\%$}                     
\def\esyscalCT {900}                           
\def\esyscalEB {250}                           
\def\esysreso  {40}                            
\def\esysai    {100}                           
\def\esysisr   {70}                            
\def\CPTdel     {0.0}                          
\def\CPTdelstat {3.0}                          
\def\CPTcalibCT {$0.2\:\%$}                    
\def\CPTcalerr  {1~MeV}                        
\def\CPTqdiff   {$5\:\%$}                      
\def\CPTrel     {0.0}                          
\def\CPTrelerr  {1.8}                          
\def\limCPT {\ensuremath{3.0\times10^{-3}}}    
\def\ntau {\ensuremath{159\,373}}              
\def\efftau {\ensuremath{89\:\%}}              
\def\purtau {\ensuremath{98\:\%}}              

\begin{document}

\begin{titlepage}
\begin{center}{\large   EUROPEAN ORGANIZATION FOR NUCLEAR RESEARCH
}\end{center}\bigskip\bigskip
\begin{flushright}
       CERN-EP-2000-056   \\ April 26, 2000
\end{flushright}
\bigskip\bigskip\bigskip\bigskip\bigskip\bigskip
\begin{center}{\huge\bf\boldmath
                         A Measurement of the {\T} Mass \vspace{4mm} \\
                         and the First CPT Test with {\T} Leptons
}\end{center}\vspace*{3cm}
\begin{center}{\LARGE The OPAL Collaboration
}\end{center}\vspace*{3cm}

\begin{center}
\begin{abstract}
\bigskip
We measure the mass of the {\tl} to be {$\mtau \pm \estat (\mathrm{stat.})
\pm \esys (\mathrm{sys.})\:\mbox{MeV}$} using {\tps} from {\Z} decays. 
To test CPT invariance we compare the masses of the positively and negatively
charged {\tls}. 
The relative mass difference is found to be smaller than {\limCPT} at the 
{$90\:\%$} confidence level.
\end{abstract}
\end{center}

\bigskip
\bigskip
\bigskip
\begin{center}{\large
(Submitted to Physics Letters B)
}\end{center}
\end{titlepage}

\thispagestyle{empty}
\vspace*{-11mm}
\begin{center}{\large        The OPAL Collaboration
}\end{center}\medskip
\begin{center}{
\small
G.\thinspace Abbiendi$^{  2}$,
K.\thinspace Ackerstaff$^{  8}$,
C.\thinspace Ainsley$^{  5}$,
P.F.\thinspace Akesson$^{  3}$,
G.\thinspace Alexander$^{ 22}$,
J.\thinspace Allison$^{ 16}$,
K.J.\thinspace Anderson$^{  9}$,
S.\thinspace Arcelli$^{ 17}$,
S.\thinspace Asai$^{ 23}$,
S.F.\thinspace Ashby$^{  1}$
% ...
}\end{center}
\pagebreak
\thispagestyle{empty}
{\small
$^{  1}$School of Physics and Astronomy, University of Birmingham,
Birmingham B15 2TT, UK
\newline
$^{  2}$Dipartimento di Fisica dell' Universit\`a di Bologna and INFN,
I-40126 Bologna, Italy
\newline
$^{  3}$Physikalisches Institut, Universit\"at Bonn,
D-53115 Bonn, Germany
\newline
% ...
$^{  5}$Cavendish Laboratory, Cambridge CB3 0HE, UK
\newline
% ...
$^{  8}$CERN, European Organization for Nuclear Research,
CH-1211 Geneva 23, Switzerland
\newline
$^{  9}$Enrico Fermi Institute and Department of Physics,
University of Chicago, Chicago IL 60637, USA
\newline
% ...
$^{ 16}$Department of Physics, Schuster Laboratory, The University,
Manchester M13 9PL, UK
\newline
$^{ 17}$Department of Physics, University of Maryland,
College Park, MD 20742, USA
\newline
% ...
$^{ 22}$Department of Physics and Astronomy, Tel Aviv University,
Tel Aviv 69978, Israel
\newline
$^{ 23}$International Centre for Elementary Particle Physics and
Department of Physics, University of Tokyo, Tokyo 113-0033, and
Kobe University, Kobe 657-8501, Japan
\newline
% ...
}
\newpage

\setcounter{page}{1}
\section{Introduction}
\label{sec:intro}

We have measured the mass of the {\tl}, {\mt}, 
using data taken by the OPAL detector during LEP running at the {\Z} resonance.
The first test of CPT invariance using {\tls} is performed by
measuring the masses of the positively and negatively charged {\tls} 
separately.
To determine the {\tm} we use two pseudomass techniques, 
previously established by ARGUS \cite{ARGUS} and CLEO \cite{CLEO},
which rely on the reconstruction of the mass, energy, and direction of 
the hadronic system in hadronic {\tds}.

In the OPAL detector charged particle tracks are reconstructed by a central
detector consisting of a silicon microvertex detector, a vertex chamber, a
large jet chamber, and z-chambers.
Photons coming from the decay of neutral pions are measured using a hermetic 
lead glass calorimeter located outside of the solenoid and an iron--streamer
tube sandwich calorimeter is used to measure hadronic showers. 
This hadronic calorimeter is used in conjunction with muon chambers mounted 
around it to separate muons from hadrons in the event selection and decay 
mode identification.
The OPAL detector is described in detail elsewhere \cite{Detector}.
\vspace*{5mm}

\section{The Pseudomass Method}
\label{sec:method}

\begin{figure}[t]
\begin{center}
\vspace*{-2mm}
\resizebox{11cm}{!}{\includegraphics{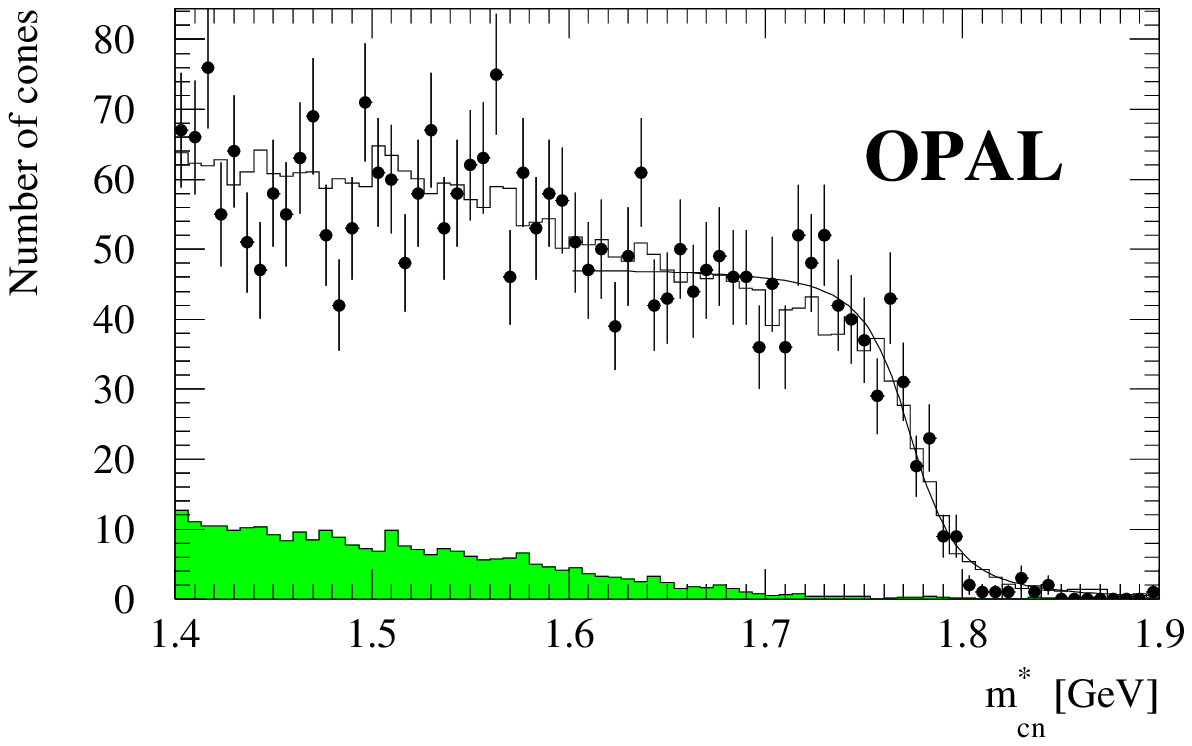}}
\vspace*{-2mm}
\end{center}
\caption{The distribution of the pseudomass {\ma} from the {\tauppp} class. 
         The points with error bars are data and the open histogram is the 
         Monte Carlo prediction including the background from misidentified
         {\tds} (shaded area) normalized to the data. 
         The solid line shows the parameterization in the fit window 
         (1.6 to 2.0 GeV) after the final unbinned maximum likelihood fit to 
         the data. 
         The events have been binned for the presentation only.} 
\label{fig:ARGUS} 
\end{figure}

In a hadronic {\td}, the mass of the {\tl} is related to the 4-momentum of 
the resulting hadronic system\footnote{Here `hadronic system'
          is defined as the sum of all the hadrons produced in the decay. The 
          sum of their momenta gives the momentum of the hadronic system.}
by the formula
\begin{equation}
  m_\nu^2 = (p_\T - \ph)^2 = m_\T^2 - 2 E_\T \Eh 
                             + 2 |\vec{p}_\T| |\vph| \cos\PSI
                             + \mh^2,
  \hspace{1cm} |\vec{p}_\T| = \sqrt{E_\T^2 - m_\T^2}
\end{equation}
where {$p_\T$} (\ph), {$\vec{p_\T}$} (\vph), and {$E_\T$} (\Eh) are the 
4-momentum, 3-momentum, and energy of the {\T} (hadronic system). 
The neutrino mass, {$m_\nu$}, is set to zero and the {\T} energy
is taken to be the beam energy,
thus permitting the mass of the {\tl} to be reconstructed if the angle {\PSI} 
between the direction of the {\T} and the hadronic system were known.
Due to the high boost at LEP energies, {\PSI} is limited to a few degrees, 
with the exact limit on {\PSI} depending on {\mh}.
A pseudomass, {\ma} (one value per cone\footnote{Cones with a half-opening 
              angle of {$35^\circ$} are defined around the leading particles
              in the event. In most cases a cone contains the decay products of
              exactly one {\T}.}), 
is defined by taking {\PSI} to be zero.
It gives the true mass when {$\PSI = 0$} and is smaller than the true mass in 
any other case. 
The distribution of {\ma} for {\tauppp} decays, presented in fig.\ 
\ref{fig:ARGUS} is a broad distribution with a sharp cutoff at the {\tm}. 
Further, the details of the shape of the distribution are determined by the 
dynamics of the decay with the general tendency that the more massive the 
hadronic system, the closer the pseudomass is to the cutoff. 
However, the position of the cutoff only depends on the mass of the {\tl}.
The {\tm} is extracted from this position.
From fig.\ \ref{fig:ARGUS} a small tail of pseudomass above the cutoff is
evident. It is caused by three effects: 
background, resolution, and initial state radiation, the latter weakens the
assumption {$E_\T = E_{\mathrm{beam}}$}.  
More details can be found in \cite{ARGUS}.

\begin{figure}[b]
\begin{center}
\vspace*{-1mm}
\resizebox{7.5cm}{!}{\includegraphics{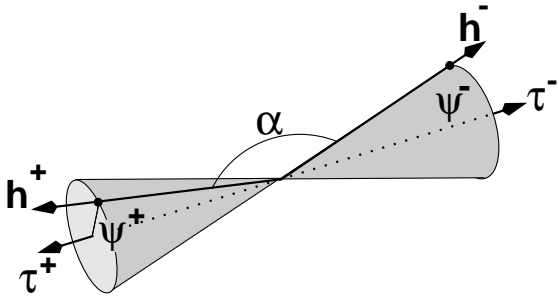}}
\end{center}
\caption{Illustration of the relevant angles: {\PSIp} and {\PSIm} are the
         opening angles between the {\tl} and its hadronic decay products. 
         {$\alpha$} is the angle between the hadrons of the two {\tls}. 
         In the absence of radiation, the {\tls} are produced collinear.}
\label{fig:angles} 
\end{figure}

\begin{figure}[t]
\begin{center}
\vspace*{-2mm}
\resizebox{9.5cm}{!}{\includegraphics{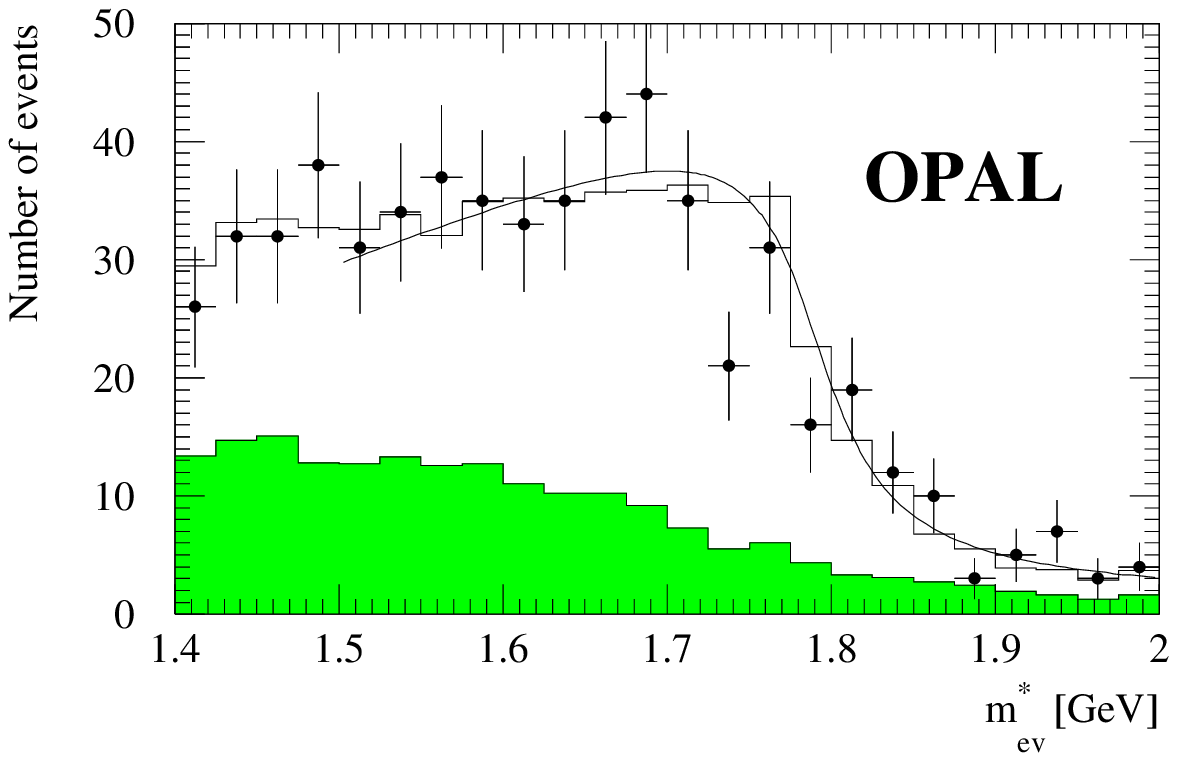}}
\vspace*{-2mm}
\end{center}
\caption{The distribution of the pseudomass {\mc} from the class of {\taup} 
         decays recoiling against \taupn. 
         The points with error bars are data and the open histogram is the 
         Monte Carlo prediction including the background from misidentified
         {\tds} (shaded area) normalized to the data. 
         The solid line shows the parameterization in the fit window 
         (1.5 to 2.0 GeV) after the final unbinned maximum likelihood fit to 
         the data. 
         The events have been binned for the presentation only.} 
\label{fig:CLEO} 
\vspace*{14mm}
\end{figure}

In {\tp} events where both {\tls} decay hadronically, a second pseudomass,
{\mc}, can be defined, which is derived from the event as a whole, not just
from one of the two cones.
In such events the acollinearity
{$\alpha$}, i.e.\ the angle between the hadronic systems from the two decays, 
carries additional information on the {\tm} (see fig.\ \ref{fig:angles}). 
This acollinearity restricts the decay angles of the {\tls} through the 
relation {$\PSIp + \PSIm + \alpha \ge \pi$}, where {\PSIp} and {\PSIm} are the 
decay angles of the positive and negative {\T} in the laboratory frame, 
respectively. 
If the sum of these three angles were known, the mass of the {\tl} could be 
reconstructed on an event by event basis. 
Since this information is unavailable the pseudomass, {\mc}, 
is calculated instead, assuming that the sum of the three angles is 180 
degrees and that the masses of the positive and negative {\T} are equal. 
The {\mc} distribution shows similar features to {\ma}. An example is shown in 
fig.\ \ref{fig:CLEO}.
Again, the cutoff at high values, from which the {\tm} is extracted, 
is clearly visible. 
More details can be found in \cite{CLEO}.
\pagebreak

\section{Selection and Classification}
\label{sec:selection}
The selection of {\tp} events is performed as described in \cite{TauSel} using
the data recorded during the years 1990 to 1995 at 
center-of-mass energies at and around the {\Z} mass. 
A total of {\ntau} {\tps} are identified within a geometrical acceptance 
limited to polar angles\footnote{We use a righthanded coordinate system with 
                   the z-axis pointing in the direction of the electron
                   beam and the x-axis pointing towards the center of LEP.}
of {$|\cos\theta| < 0.95$}.
The overall efficiency including the geometrical acceptance is {\efftau}. 
The purity of the sample is {\purtau}. The background consists mainly of 
electron and muon pairs.

Each selected event is split into two narrow cones of particles, 
one for each {\T}. 
All 1-prong and 3-prong cones\footnote{A
               1- or 3-prong cone is a cone containing 1 or 3 charged tracks.}
with a minimum momentum of {$5\:\%$} of the beam momentum are then subjected 
to a likelihood classification in order to identify the decay mode of the two 
{\tls} using the same procedure as in \cite{TauID}. 
Table \ref{tab:likelihood} summarizes the decay modes considered. 

\begin{table}[tb]
  \begin{center}
    \vspace*{4mm}
    \begin{tabular}{|l|r@{~$\%$~}|r@{~$\%$~}|r|
                    |l|r@{~$\%$~}|r@{~$\%$~}|r|}
    \hline
    \multicolumn{4}{|c||}{1-prong modes} &
    \multicolumn{4}{|c|} {3-prong modes} \\
    \hline
             &eff.\ &pur.\ & decays &         &eff.\ &pur.\ & decays \\
    \hline \hline
     \taup   & 76.2 & 64.5 &  41148 & \tauppp \rule{0pt}{4mm}
                                              & 71.5 & 83.8 &  24832 \\
     \taupn  & 60.2 & 77.1 &  58518 & \taupppn& 61.6 & 64.2 &  11636 \\
     \taupnn & 45.1 & 40.9 &  26093 &         & \emp & \emp &        \\
    \hline
    \end{tabular}
    \caption{Efficiency, purity, and number of decays selected by the 
             likelihood identification in the decay modes explored in this
             analysis. In addition to these modes, the likelihood identifies 
             leptonic decays and various modes with one charged hadron plus 
             additional tracks from photon conversions or Dalitz decays.
             All numbers refer to the full range of pseudomasses. 
             The purity in the region of the cutoff is substantially better.}
    \label{tab:likelihood}
  \end{center}
\end{table}

For the calculation of the pseudomasses {\ma} and {\mc}, all tracks are 
assumed to originate from charged pions. 
Neutral pions are reconstructed from showers in the electromagnetic calorimeter
with the algorithm described in \cite{TauID}.
The number of neutral pions reconstructed in a cone does not necessarily match
the number of {\pni} expected from the decay mode identified by the likelihood
selection. 
For example, there might only be a single {\pni} reconstructed in a cone
identified as a {\taupnn} decay.
In such a case the missing neutral pions are ignored in the calculation of the
4-momentum of the hadronic system.
If there are too many {\pni} reconstructed, we exclude those with the lowest 
energies.

\begin{table}[tb]
 \begin{center}
   \vspace*{4mm}
   \begin{tabular}{|l||c|c|c|c|c|}

    \hline       \raisebox{-1mm}[5mm][0pt]
                 & \ppi & \ppi\pni & \hspace{-1mm}\ppi2\pni\hspace{-1mm} & 
                   3 \ppi & \hspace{-2mm} 3\ppi \pni \hspace{-1mm} \\
    \hline \hline  
     no ID       & {\bf --}  & \hspace{-2.8mm}
                               \raisebox{-3mm}{
                                     \setlength{\unitlength}{1mm}
                                     \begin{picture}(9,9)
                                     \put(0.,1.){{\bf --}}
                                     \put(5.0,5.1){cn}
                                     \put(10,0){\line(-3,2){13.5}}
                                     \end{picture}} 
                               \hspace{-2.5mm} & 
                               \hspace{-2.8mm}
                               \raisebox{-3mm}{
                                     \setlength{\unitlength}{1mm}
                                     \begin{picture}(9,9)
                                     \put(0.,1.){{\bf --}}
                                     \put(5.0,5.1){cn}
                                     \put(10,0){\line(-3,2){13.5}}
                                     \end{picture}} 
                               \hspace{-2.5mm} & 
                               \hspace{-2.8mm}
                               \raisebox{-3mm}{
                                     \setlength{\unitlength}{1mm}
                                     \begin{picture}(9,9)
                                     \put(0.,1.){{\bf --}}
                                     \put(5.0,5.1){cn}
                                     \put(10,0){\line(-3,2){13.5}}
                                     \end{picture}} 
                               \hspace{-2.5mm} & 
                               \hspace{-2.8mm}
                               \raisebox{-3mm}{
                                     \setlength{\unitlength}{1mm}
                                     \begin{picture}(9,9)
                                     \put(0.,1.){{\bf --}}
                                     \put(5.0,5.1){cn}
                                     \put(10,0){\line(-3,2){13.5}}
                                     \end{picture}} 
                               \hspace{-2.5mm} \\
    \hline
     \ppi        & ev    & ev        & ev           & 
                               \hspace{-2.8mm}
                               \raisebox{-3mm}{
                                     \setlength{\unitlength}{1mm}
                                     \begin{picture}(9,9)
                                     \put(0.,1.){{\bf --}}
                                     \put(5.0,5.1){cn}
                                     \put(10,0){\line(-3,2){13.5}}
                                     \end{picture}} 
                               \hspace{-2.5mm} & 
                               \hspace{-2.8mm}
                               \raisebox{-3mm}{
                                     \setlength{\unitlength}{1mm}
                                     \begin{picture}(9,9)
                                     \put(0.,1.){{\bf --}}
                                     \put(5.0,5.1){cn}
                                     \put(10,0){\line(-3,2){13.5}}
                                     \end{picture}} 
                               \hspace{-2.5mm} \\
    \hline
     \ppi\pni    &      & ev        & 
                               \hspace{-2.8mm}
                               \raisebox{-3mm}{
                                     \setlength{\unitlength}{1mm}
                                     \begin{picture}(9,9)
                                     \put(-0.5,1.){cn}
                                     \put(5.0,5.1){cn}
                                     \put(10,0){\line(-3,2){13.5}}
                                     \end{picture}} 
                               \hspace{-2.5mm} & 
                               \hspace{-2.8mm}
                               \raisebox{-3mm}{
                                     \setlength{\unitlength}{1mm}
                                     \begin{picture}(9,9)
                                     \put(-0.5,1.){cn}
                                     \put(5.0,5.1){cn}
                                     \put(10,0){\line(-3,2){13.5}}
                                     \end{picture}} 
                               \hspace{-2.5mm} & 
                               ev \\
    \hline
     \ppi\:2\,\pni &      &          & 
                               \hspace{-2.8mm}
                               \raisebox{-3mm}{
                                     \setlength{\unitlength}{1mm}
                                     \begin{picture}(9,9)
                                     \put(-0.5,1.){cn}
                                     \put(5.0,5.1){cn}
                                     \put(10,0){\line(-3,2){13.5}}
                                     \end{picture}} 
                               \hspace{-2.5mm} & 
                               \hspace{-2.8mm}
                               \raisebox{-3mm}{
                                     \setlength{\unitlength}{1mm}
                                     \begin{picture}(9,9)
                                     \put(-0.5,1.){cn}
                                     \put(5.0,5.1){cn}
                                     \put(10,0){\line(-3,2){13.5}}
                                     \end{picture}} 
                               \hspace{-2.5mm} & 
                               \hspace{-2.8mm}
                               \raisebox{-3mm}{
                                     \setlength{\unitlength}{1mm}
                                     \begin{picture}(9,9)
                                     \put(-0.5,1.){cn}
                                     \put(5.0,5.1){cn}
                                     \put(10,0){\line(-3,2){13.5}}
                                     \end{picture}} 
                               \hspace{-2.5mm} \\
    \hline
     3 \ppi      &      &          &             & 
                                                              \hspace{-2.8mm}
                               \raisebox{-3mm}{
                                     \setlength{\unitlength}{1mm}
                                     \begin{picture}(9,9)
                                     \put(-0.5,1.){cn}
                                     \put(5.0,5.1){cn}
                                     \put(10,0){\line(-3,2){13.5}}
                                     \end{picture}} 
                               \hspace{-2.5mm} & 
                               \hspace{-2.8mm}
                               \raisebox{-3mm}{
                                     \setlength{\unitlength}{1mm}
                                     \begin{picture}(9,9)
                                     \put(-0.5,1.){cn}
                                     \put(5.0,5.1){cn}
                                     \put(10,0){\line(-3,2){13.5}}
                                     \end{picture}} 
                               \hspace{-2.5mm} \\
    \hline
     3 \ppi \pni & \hspace*{9mm} &          &             &       & 
                               \hspace{-2.8mm}
                               \raisebox{-3mm}{
                                     \setlength{\unitlength}{1mm}
                                     \begin{picture}(9,9)
                                     \put(-0.5,1.){cn}
                                     \put(5.0,5.1){cn}
                                     \put(10,0){\line(-3,2){13.5}}
                                     \end{picture}} 
                               \hspace{-2.5mm} \\
    \hline
   \end{tabular}
   \caption{The assignment of the various decay modes from the likelihood 
            identification into classes analyzed by either {\ma} (cn) or 
            {\mc} (ev).
            The row labeled `no ID' combines the cones where no identification
            was possible with those identified as leptonic decays and all 
            other channels not used in the analysis. {\tauppp} includes
            {\taupk}. For {\taup} versus {\tauppp} see text.}
   \label{tab:ass}
 \end{center}
 \vspace*{4mm}
\end{table}

The pseudomass used to analyze an event depends on its identified decay modes.
This choice of method can either be {\ma} applied to one or both cones of the 
event independently or {\mc} applied to the whole event. 
Table \ref{tab:ass} illustrates the association of the decay modes of the {\T}
to the two methods. 
The association has been optimized to give the smallest possible statistical 
error on {\mt},
a process which results in decays with high hadronic masses typically being 
assigned to the {\ma} method, leaving the lighter hadronic masses for {\mc}.
Every decay or event is only analyzed by one method to avoid statistical
correlations.

A slightly more complicated procedure is used for events with a {\tauppp}
decay recoiling against a {\taup}. 
The 3-prong cone is initially subjected to the {\ma} method and the 1-prong 
cone neglected (see Tab.\ \ref{tab:ass}). 
If {\ma} turns out to be below the fit window described in the following 
section, the event is analyzed through {\mc} instead.
This procedure could also be applied to other combinations of decay modes, 
but the gain in overall resolution is only marginal.

In the following steps of the analysis the distributions of {\ma} 
are handled in exactly the same way as those of {\mc}.
We define the following classes (see Tab.\ \ref{tab:ass}):
\begin{itemize}
\item Four classes corresponding to the decay modes {\taupn}, {\taupnn}, 
      {\tauppp}, and {\taupppn} which are analyzed with the {\ma} method;
\item Five classes for the different combinations of decay modes
      analyzed with the {\mc} method;
\item One class for the {\taup} versus {\tauppp} decays
      described in the previous paragraph.
\end{itemize}

Each class is further subdivided into subclasses according to the 
expected resolution. 
This avoids having information from well reconstructed entries degraded
by the overlay of those entries with poorer resolution.
There is a total of 25 subclasses, between one and five per class.
The criteria used for the classification of the expected resolution is the
quality of the reconstruction of the hadronic system, features like the 
presence of z-chamber or silicon hits on the tracks, the number and momentum 
of the reconstructed neutral pions, the probability of a vertex fit or the 
polar angle of the cone.
The best resolution\footnote{The resolution is defined as the r.m.s of the
                             difference between true and reconstructed 
                             pseudomass from simulated events that fall into 
                             the fit window.}
in the pseudomass {\ma} of 17~MeV is achieved in the {\tauppp} class,
and this class also has the largest weight in the final result.
For {\mc}, the best resolution is 21~MeV in the {\taup} versus {\taup} class.
Table \ref{tab:classes} gives some more information on the classes.
Figures \ref{fig:ARGUS} and \ref{fig:CLEO} correspond to the subclass with the
best resolution.

\begin{table}[t]
  \begin{center}
    \begin{tabular}{|l|c|r|r@{$\;\pm\!\!$}r|}
      \hline
      class & method & entries & \multicolumn{2}{c|}{result in MeV} \\
      \hline
      \ppi   \pni \rule{0pt}{4mm} & \ma &  1467 & 1777.6  & 7.2  \\
      \ppi 2 \pni                 & \ma &  1311 & 1801.7  & 12.6  \\
      3 \ppi                      & \ma &  2680 & 1776.1  & 1.9  \\
      3 \ppi \pni                 & \ma &  3467 & 1777.6  & 6.2  \\
      \hline
      \ppi \dd \ppi \rule{0pt}{4mm} & \mc &   483 & 1790.2  &  9.4  \\
      \ppi \dd \ppi\pni             & \mc &  1346 & 1759.9  &  6.4  \\
      \ppi \dd \ppi \:2\,\pni       & \mc &   506 & 1779.9  & 16.4  \\
      \ppi\pni \dd \ppi\pni         & \mc &  1191 & 1770.6  &  8.7  \\
      \ppi\pni \dd 3 \ppi\pni       & \mc &   565 & 1793.3  & 13.5  \\
      \hline
      \ppi \dd 3 \ppi \rule{0pt}{4mm} & \mc &   328 & 1757.9  & 10.9  \\
      \hline
    \end{tabular}
    \caption{The 10 classes of the analysis. The method applied, number of
             entries in the fit window, and a result (statistical error only) 
             from a fit to each class alone are specified. The 
             `\ppi \dd 3 \ppi' class only contains those events not used 
             in the `3 \ppi' class (see text).}
    \label{tab:classes}
  \end{center}
\end{table}
\pagebreak

\section{Extraction of the {\boldmath \T} Mass}
\label{sec:result}

The value of the {\tm} is extracted from the pseudomass distribution by 
means of unbinned maximum likelihood fits. 
The distributions are parameterized by a step function 
multiplied by a polynomial of up to third order. 
For each subclass one of the following step functions is chosen, 
depending on which gives the best description of the data: 
\begin{equation}
  f\left(x\right) = \left\{
  \begin{array}{l}
  \displaystyle \frac{1}{2} \left(1 - \frac{x}{\sqrt{1 + x^2}}\right) 
                                               {\rule[-7mm]{0pt}{10mm}} \\
  \displaystyle \frac{1}{1 + e^x}               
                                               {\rule[-7mm]{0pt}{10mm}} \\
  \displaystyle \frac{1}{\pi} \left(\frac{\pi}{2} - \arctan\left(x\right)
                                                              \right)
  \end{array}\right.
\end{equation}
The variable {$x$} is related to the pseudomass by
{$x = b_i\left(a_i + \delta + m^\ast\right)$} 
where {$a_i$} and {$b_i$} are fit parameters and {$m^\ast$} is the 
pseudomass, the measured quantity.
The index {$i$} labels the subclasses.
The parameter {$\delta$} is a common shift of all distributions.
The order of the polynomial is increased from zero until a reasonably
good parameterization of the distribution is achieved.\footnote{We histogram 
                       the data and apply a {$\chi^2$} test to decide on the 
                       quality of the description.}
Only events in the vicinity of the cutoff are used for the fits. 
The upper end of the fitting window is 2 GeV for all subclasses.  
The values of the lower bound range from 1.4 to 1.6 GeV, depending on the 
subclass.
For each subclass the lower bound is decreased from 1.6 GeV until no further
significant reduction of the expected statistical error is observed. 
After the initial parameterization of the {\mstar} distributions, the fit 
windows are reduced by 50~MeV on each side
to ensure a stable parameterization in all areas that might be reached in 
the course of the following fits.
The type of step function, the order of the polynomial, and the size of the 
fitting window are determined for each subclass separately from a Monte Carlo 
simulation \cite{MonteCarlo}.

The functional dependence of the pseudomass distribution on {\mt} is 
very simple in the vicinity of the cutoff. 
In this region the shape of the distribution does not depend on {\mt}.
On a variation of {\mt} the whole distribution is shifted along the pseudomass
axis by an amount identical to the shift in {\mt}.
We will use this relation to extract the result, although it is not strictly 
correct away from the cutoff.
We have checked that the fit windows are small enough so as not to introduce a 
systematic bias from this procedure.

The {\tm} is extracted in three steps. In the first step, the distributions 
from simulated events are fitted for each subclass separately in order to fix 
the parameters of the fit function, i.e. the position and width of the step
function ($a_i, b_i$), and the coefficients of the polynomials. 
In the second step the same simulated distributions are simultaneously fitted 
using the slightly reduced fit window.
A common shift {$\delta_{\mathrm{MC}}$} is the only free parameter in this fit
with all other parameters fixed to the values obtained in the first step. 
In the third step, the same common fit is applied to the data, but with 
{$\delta_{\mathrm{data}}$} as the free parameter. 
The result is calculated from {$m_\tau = m_\tau^{\mathrm{MC}} + 
\delta_{\mathrm{data}} - \delta_{\mathrm{MC}}$} with {$m_\tau^{\mathrm{MC}} = 
1777\:\mbox{MeV}$} the {\tm} used in the simulation.
The result is {\mtau}~MeV with a statistical error of {\estatMC}~MeV from the 
Monte Carlo and {\estat}~MeV from the data.
Results were also determined for each class separately as a cross check, 
the results are shown in table \ref{tab:classes}.

\section{Systematic Uncertainties }
\label{sec:systematics}

The most significant contributions to the overall systematic uncertainty 
are described in this section.
\begin{itemize}
\item The dominant systematic uncertainty comes from the calibration of the
      tracking chambers and the electromagnetic calorimeter.
      We use muons and electrons of 45~GeV from {\Z} decays and Bhabha 
      scattering to check these calibrations.
      We find uncertainties in the momentum scale of the tracking chambers 
      relative to the Monte Carlo of {\calibCT} and for the calorimeter of 
      less than {\calibEB}. 
      Several different scenarios are employed to extrapolate these 
      uncertainties to the momentum and energy ranges relevant to {\tds}.
      The most pessimistic scenario, a scaling of {$1/p_T$} by a factor,
      gives a systematic error from the tracking of {\esyscalCT}~keV. 
      The uncertainty in the calibration of the calorimeter is found to have 
      a negligible impact ({\esyscalEB}~keV), since it only effects channels 
      with neutral pions.
\item Uncertainties in modeling the resolution of the tracking chambers
      result in  an error of {\esysreso}~keV on the measurement of {\mt}. 
      The resolution of the electromagnetic calorimeter has a negligible 
      impact.   
\item In order to estimate the systematic error contribution from the modeling
      of the dynamics of the {\tds}, we have varied the mass and width of the 
      {$\mathrm{a}_1$} meson in the {$3\ppi$} final state by 
      {$\pm 100\:\mbox{MeV}$} and
      changed the fraction of {$\omega\pi$} events in the {$4\pi$} final state
      by {$\pm 20\:\%$}.
      The first two variations each change the result by {\esysai}~keV, while
      the third has a negligible impact.
      We also tested two different models for the {\tauppp} decay \cite{Model}.
      Using these models the result changes by less than {\esysai}~keV.
\item In the derivation of the pseudomass formulae the assumption {$E_\tau = 
      E_{\mathrm{beam}}$} is used, which is only true in the absence of 
      initial state radiation. 
      Low energy photons from initial state radiation have only a negligible 
      effect on the measurement, while a variation of {$\pm 1\:\%$} of the 
      rate of hard initial state radiation results in an uncertainty
      on the measurement of {\esysisr}~keV.
\item The uncertainty in the calibration of the beam energy of LEP has a 
      negligible effect. An error of 2~MeV on the beam energy changes the
      result by less than $\pm\esysbeam$~keV.
\item The calculation of the pseudomasses assumes that the {\nut} has zero 
      mass. 
      A value of \esysmnu~MeV, the current limit on the {\T} neutrino mass
      \cite{PDG}, increases the result by 110~keV.
      No systematic error is assigned.
\item Systematic uncertainties related to background from misidentified 
      {$\tau$} decays are negligible, since their pseudomass distributions 
      show no significant structure in the region of the cutoff.
      Background from non-$\tau$ sources is also negligible. 
\item To address possible biases from the parameterization of the pseudomass
      distributions, 
      we increased the lower edge of the window for all channels, in steps of
      25~MeV, to 1.6~GeV. Furthermore, we fitted the distributions with the 
      type of step function giving the second best description of the 
      data.
      Both studies showed consistent results.
\end{itemize}
Adding in quadrature all systematic uncertainties including the statistical 
error from the Monte Carlo gives a total systematic uncertainty on the 
measurement of \esys~MeV.

\section{CPT Test}
\label{sec:CPT}

To test CPT invariance we compare the masses of the positively and negatively
charged {\tls}.
The pseudomass method {\ma} allows a separate measurement for the 
positive and the negative {\tl}, whereas the {\mc} method implicitly assumes 
the two masses to be identical and therefore the {\ma} method is used for all
decays.
The measurement does not use any Monte Carlo simulation which 
reduces the systematic uncertainties.
 
For the CPT analysis the {\ma} distributions are separated according to the 
net charges of the cones.
To extract the result we use a procedure similar to the one used to measure
{\mt}. 
In the first step, the distributions of the positive or negative {\tls}  
are fitted subclass by subclass to fix the parameterizations. 
Then all the distributions of the positive {\tls} are fitted simultaneously 
with the parameters fixed to the values from the first fit, allowing only for 
a common shift in mass {$\delta^+$}. 
Then negative {\tls} are fitted the same way with {$\delta^-$} as the free 
parameter. 
If CPT invariance is conserved, the two shifts must be equal. 
We obtain the result 
{$\delta^+ - \delta^- = \CPTdel \pm \CPTdelstat \: \mbox{MeV}$}.

In a {\tp} event produced in {\ee} collisions, a mass difference between the 
positive and negative {$\tau$} will also create a difference in energy between
the two {\T}.
Although in principle this invalidates the assumption 
{$E_\tau = E_{\mathrm{beam}}$}, this effect is numerically negligible.

Most sources of systematic errors affect the result for the positive and 
negative lepton in the same way, so that their contributions cancel. 
The largest error in the mass difference 
comes from possible differences in the calibration between positively and 
negatively charged tracks which is limited to less than {\CPTcalibCT} by 
studying muon pairs. 
This implies a systematic uncertainty on the mass difference of {\CPTcalerr}. 
All other systematic errors are negligible and the total systematic error 
is small compared to the statistical error. 
It is added in quadrature to the statistical error.

The {\ma} method uses the momentum and
mass of the hadronic system measured from the curvature of the 
tracks in the magnetic field of the detector. These tracks originate mainly
from pions with a small fraction coming from kaons. The charge and the charge 
to mass ratio of these particles have to be known in order to convert the 
observed curvature into a momentum measurement. 
However, without assuming CPT invariance, it is no longer obvious that the 
charge and mass are the same for positively and negatively charged 
pions\footnote{A common 
                  deviation of the charge of positive and negative pion 
                  from the charge of the electron or the {$\tau$} would
                  not spoil the measurement. Only a difference between 
                  {$\pi^+$} and {$\pi^-$} is relevant.}.
Experimentally, the relative pion mass difference has been measured to be 
{$(2 \pm 5) \times 10^{-4}$} which is below our sensitivity, but the
pion charge difference has not been measured separately.
However, the pion charges could be different due to a charge difference between
the {$\tau^+$} and the {$\tau^-$} or because of charge non-conservation.
A charge difference between {$\pi^+$} and {$\pi^-$} as large as  
{\CPTqdiff} would invalidate our measurement.

This is the first test of CPT invariance with {\tls}. The result is
\vspace*{-1.5mm}
\begin{equation}
  \frac{m{\scriptstyle(}\tau^{\scriptscriptstyle+}{\scriptstyle)}
      - m{\scriptstyle(}\tau^{\scriptscriptstyle-}{\scriptstyle)}}
       {m{\scriptstyle(} \tau{\scriptstyle )}}
  = \left(\CPTrel \pm \CPTrelerr\right) \times 10^{-3}.
  \label{eq:CPTresult}
\end{equation}
\vspace*{-5mm}\\
where {$m(\tau)$} is the charge-independent result from the previous section.
This result implies that the relative {\T} mass difference is smaller than 
{\limCPT} at the {$90\:\%$} confidence level.

\section{Summary}
\label{Summary}

\begin{figure}[hb]
  \begin{center}
    \resizebox{10cm}{!}{\includegraphics{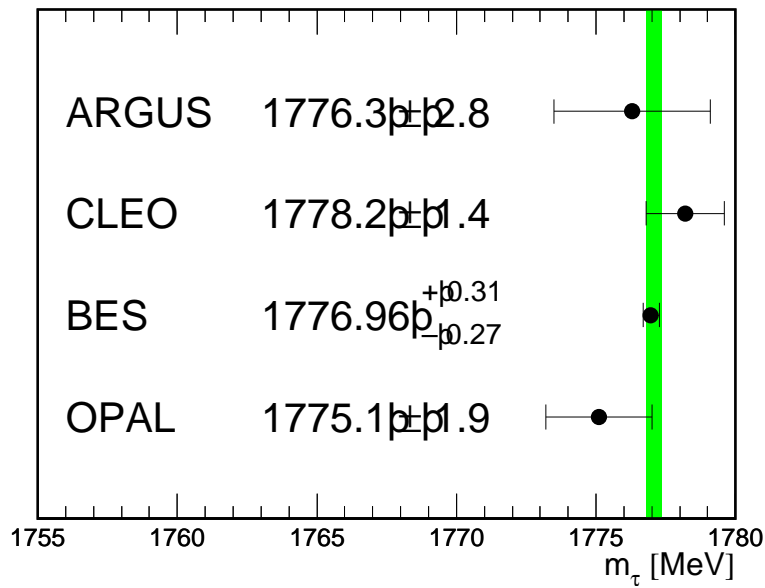}}
    \caption{Comparison of our result with the results from 
             ARGUS \cite{ARGUS}, CLEO \cite{CLEOres}, and BES \cite{BESres}.
             The shaded band indicates the current world average \cite{PDG},
             which does not include the OPAL result.}
    \label{fig:comparison}
  \end{center}
\end{figure}

We have measured the mass of the {\tl} from the pseudomass distributions of the
hadronic {\tds} recorded by the OPAL detector at LEP and obtained a result of  
\begin{equation}
  \mt = \mtau \pm \estat (\mathrm{stat.}) \pm \esys (\mathrm{sys.})\;\mbox{MeV}
  \label{eq:result}
\end{equation}
where the first error is statistical and the second systematic. 
The result is compared to other measurements in fig.\ \ref{fig:comparison}.
Using the pseudomass {\ma} we obtain an independent measurement of the mass
of the positive and negative {\T}. The two numbers are consistent and a limit 
of {\limCPT} on the relative mass difference is be placed at the {$90\:\%$}
confidence level. This is the first test of CPT invariance in {\T} physics.

\section*{Acknowledgments}
We particularly wish to thank the SL Division for the efficient operation
of the LEP accelerator at all energies
 and for their continuing close cooperation with
our experimental group.  We thank our colleagues from CEA, DAPNIA/SPP,
CE-Saclay for their efforts over the years on the time-of-flight and trigger
systems which we continue to use.  In addition to the support staff at our own
institutions we are pleased to acknowledge the  \\
Department of Energy, USA, \\
National Science Foundation, USA, \\
Particle Physics and Astronomy Research Council, UK, \\
Natural Sciences and Engineering Research Council, Canada, \\
Israel Science Foundation, administered by the Israel
Academy of Science and Humanities, \\
Minerva Gesellschaft, \\
Benoziyo Center for High Energy Physics,\\
Japanese Ministry of Education, Science and Culture (the
Monbusho) and a grant under the Monbusho International
Science Research Program,\\
Japanese Society for the Promotion of Science (JSPS),\\
German Israeli Bi-national Science Foundation (GIF), \\
Bundesministerium f\"ur Bildung und Forschung,
Forschung und Technologie, Germany, \\
National Research Council of Canada, \\
Research Corporation, USA,\\
Hungarian Foundation for Scientific Research, OTKA T-029328, 
T023793 and OTKA F-023259.\\



\begin{thebibliography}{99}
\bibitem{ARGUS}
   ARGUS Collaboration, H.~Albrecht \etal,
   Phys.\ Lett.\ {\bf B 292} (1992) 221.
\bibitem{CLEO}
   CLEO Collaboration, R.~Ballest \etal,
   Phys.\ Rev.\ {\bf D 47} (1993) 3671.
\bibitem{Detector} 
   OPAL Collaboration, K.~Ahmet \etal,
   Nucl.\ Instr.\ and Meth.\ {\bf A 305} (1991) 275; \\
   P.P.~Allport \etal, 
   Nucl.\ Instr.\ and Meth.\ {\bf A 324} (1993) 34; \\
   P.P.~Allport \etal, 
   Nucl.\ Instr.\ and Meth.\ {\bf A 346} (1994) 476.
\bibitem{TauSel}
   OPAL Collaboration, G.~Alexander \etal,
   Phys.\ Lett.\ {\bf B 266} (1991) 201,\\
   OPAL Collaboration, R.~Akers \etal,
   Phys.\ Lett.\ {\bf B 328} (1994) 207.
\bibitem{TauID}
   OPAL Collaboration, J.~Allison \etal,
   Z.\ Phys.\ {\bf C 66} (1995) 31;\\
   OPAL Collaboration, K.~Ackerstaff \etal,
   Eur.\ Phys.\ J.\ {\bf C 7} (1999) 571.
\bibitem{MonteCarlo}
   KORALZ, Version 4.02,
   S.~Jadach, B.~F.~L.~Ward, and Z.~W\c{a}s,
   Comp.\ Phys.\ Comm.\ {\bf 79} (1994) 503.\\
   TAUOLA, Version 2.5,
   S.~Jadach, Z.~W\c{a}s, R.~Decker, and J.~H.~K\"uhn,
   Comp.\ Phys.\ Comm.\ {\bf 76} (1993) 361.
\bibitem{Model}
   J.~H.~K\"uhn and A.~Santamaria, Z.\ Phys.\  {\bf C 48} (1990) 445; \\
   N.~Isgur, C.~Morningstar and C.~Reader, Phys.\ Rev.\  {\bf D 39}
   (1989) 1357.
\bibitem{PDG}
   Particle Data Group,
   C.~Caso \etal,
   Euro.\ Phys.\ J.\ {\bf C 3} (1998) 1.
\bibitem{CLEOres}
   CLEO Collaboration, A.~Anastassov \etal,
   Phys.\ Rev.\ {\bf D 55} (1997) 2559, ibid.\ {\bf D 58} (1998) 119904.
\bibitem{BESres}
   BES Collaboration, J.Z.~Bai \etal,
   Phys.\ Rev.\ {\bf D 53} (1996) 20.

\end{thebibliography}
\end{document}